\documentclass[useAMS,usenatbib]{mn2e}
\usepackage{graphicx}
\title[The most metal-poor stars in the Galactic bulge]{The Gaia-ESO Survey: the most metal-poor stars in the Galactic bulge}
\author[L. M. Howes et al.]{L. M. Howes,$^{1}$\thanks{E-mail: louise.howes@anu.edu.au} M. Asplund,$^{1}$ A. R. Casey,$^{2}$ S. C. Keller,$^{1}$ D. Yong,$^{1}$  G. Gilmore,$^{2}$ \newauthor 
K. Lind,$^{2,3}$ C. Worley,$^{2}$ M. S. Bessell,$^{1}$ L. Casagrande,$^{1}$ A. F. Marino,$^{1}$ D. M. Nataf,$^{1}$ \newauthor
C. I. Owen,$^{1}$ G. S. Da Costa,$^{1}$ B. P. Schmidt,$^{1}$ P. Tisserand,$^{1}$ S. Randich,$^{4}$ S. Feltzing,$^{5}$ \newauthor
A. Vallenari,$^{6}$ C. Allende Prieto,$^{7,8}$  T. Bensby,$^{5}$ E. Flaccomio,$^{9}$ A. J. Korn,$^{3}$ E. Pancino,$^{10,11}$ \newauthor
A. Recio-Blanco,$^{12}$ R. Smiljanic,$^{13}$ M. Bergemann,$^{2}$ M. T. Costado,$^{14}$ F. Damiani,$^{9}$  \newauthor
U. Heiter,$^{3}$ V. Hill,$^{12}$ A. Hourihane,$^{2}$ P. Jofr\'e,$^{2}$  C. Lardo,$^{10}$ P. de Laverny,$^{12}$ L. Magrini,$^{4}$ \newauthor
E. Maiorca,$^{4}$ T. Masseron,$^{2}$ L. Morbidelli,$^{4}$ G. G. Sacco,$^{4}$ D. Minniti,$^{15,16,17}$ M. Zoccali$^{17,18}$
\\
$^{1}$Research School of Astronomy and Astrophysics, Australian National University, ACT 2601, Australia\\
$^{2}$Institute of Astronomy, University of Cambridge, Madingley Road, Cambridge, CB3 0HA, United Kingdom\\
$^{3}$Department of Physics and Astronomy, Division of Astronomy and Space Physics, Uppsala University, Box 516, SE-751 20 Uppsala, Sweden\\
$^{4}$INAF - Osservatorio Astrofisico di Arcetri, Largo E. Fermi 5, 50125, Florence, Italy\\
$^{5}$Lund Observatory, Department of Astronomy and Theoretical Physics, Box 43, SE-221 00 Lund, Sweden\\
$^{6}$INAF - Padova Observatory, Vicolo dell'Osservatorio 5, 35122 Padova, Italy\\
$^{7}$Instituto de Astrofisica de Canarias, E-38205 La Laguna, Tenerife, Spain\\
$^{8}$Departamento de Astrofisica, Universidad de La Laguna, E-38206 La Laguna, Tenerife, Spain\\
$^{9}$INAF - Osservatorio Astronomico di Palermo, Piazza del Parlamento 1, 90134, Palermo, Italy\\
$^{10}$INAF - Osservatorio Astronomico di Bologna, via Ranzani 1, 40123, Bologna, Italy\\
$^{11}$ASI Science Data Center, Via del Politecnico SNC, 00133 Roma, Italy\\
$^{12}$Laboratoire Lagrange (UMR7293), Universit\'e de Nice Sophia Antipolis, CNRS, Observatoire de la C\^ote d'Azur, CS 34229, \\F-06304 Nice cedex 4, France\\
$^{13}$Department for Astrophysics, Nicolaus Copernicus Astronomical Center, ul. Rabia\'{n}ska 8, 87-100 Toru\'{n}, Poland\\
$^{14}$Instituto de Astrof\'{i}sica de Andaluc\'{i}a-CSIC, Apdo. 3004, 18080 Granada, Spain\\
$^{15}$Departamento de Ciencias Fisicas, Universidad Andres Bello, Republica 220, Santiago, Chile\\
$^{16}$Vatican Observatory, V00120 Vatican City State, Italy\\
$^{17}$Millenium Institute of Astrophysics, Av. Vicu\~na Mackenna 4680, Macul, Santiago, Chile\\
$^{18}$Instituto de Astrof\'{i}sica, Facultad de F\'{i}sica, Pontificia Universidad Cat\'{o}lica de Chile, Av. Vicu\~{n}a Mackenna, Santiago, Chile}

\begin{document}

\date{Accepted 2014 September 23.  Received 2014 September 20; in original form 2014 July 14}

\pagerange{\pageref{firstpage}--\pageref{lastpage}} \pubyear{2014}

\maketitle

\label{firstpage}

\begin{abstract}
We present the first results of the EMBLA survey (Extremely Metal-poor BuLge stars with AAOmega), aimed at finding metal-poor stars in the Milky Way bulge, where the oldest stars should now preferentially reside.  EMBLA utilises SkyMapper photometry to pre-select metal-poor candidates, which are subsequently confirmed using AAOmega spectroscopy. We describe the discovery and analysis of four bulge giants with $-2.72$$\le$[Fe/H]$\le$$-2.48$, the lowest metallicity bulge stars studied with high-resolution spectroscopy to date. Using FLAMES/UVES spectra through the Gaia-ESO Survey we have derived abundances of twelve elements.  Given the uncertainties, we find a chemical similarity between these bulge stars and halo stars of the same metallicity, although the abundance scatter may be larger, with some of the stars showing unusual [$\alpha$/Fe] ratios.
\end{abstract}

\begin{keywords}
Galaxy: bulge; Galaxy: evolution; stars: abundance; stars: Population II
\end{keywords}

\section{Introduction}

The first stars in the Universe (referred to as Population III stars) have been extensively searched for, both in the local Universe (\citealt{2013pss5.book...55F} and references therein) and at high redshift (e.g., \citealt{2006ApJ...642..382B}, \citealt{2012MNRAS.425..347C}), but despite massive efforts no true Population III star has yet been found. There is an argument that no such stars should remain today: models of their formation indicate that they would have been massive and short-lived (\citealt{2001ApJ...548...19N}; \citealt{2002Sci...295...93A}). Recent simulations however have suggested that disc fragmentation could have produced smaller mass stars, some of which may have survived to the present day \citep{2011Sci...331.1040C}.
\\
Surveys focusing on the discovery of these old and metal-poor stars have almost exclusively targeted the Galactic halo (e.g., \citealt{2008A&A...484..721C}), although some more recent studies have looked at dwarf galaxies of the Local Group (e.g., \citealt{2010Natur.464...72F}). The halo is known to be on average more metal-poor than other Galactic components, and some of these halo stars pass through the solar neighbourhood, making them relatively uncomplicated to observe. The number of metal-poor halo stars discovered has been growing, and there are now chemical abundances for $>400$ metal-poor stars with [Fe/H]$<-2.5$ \citep{2013ApJ...762...25N}.
\\
It is not obvious though that the halo is the ideal place to look for the first stars. Using $\Lambda$CDM simulations, \citet{2005MNRAS.364..367D} predicted that if any were to survive to the present day, 30-60\% of them would reside within the inner 3 kpc of the Galaxy, a population density of first stars that would be 1000 times greater than that of the solar neighbourhood. \citet{2010ApJ...708.1398T} has modelled the current spatial distribution of all stars with [Fe/H]$<-3.5$ that formed prior to $z=15$,  and shown that, because of the inside-out construction of dark-matter haloes, the oldest as well as the most metal-poor stars should be more frequent in the central regions of the Galaxy. In other words, even if these stars may have originated elsewhere, they are now most likely to be located within the central regions of the Galaxy.
\\
Few, if any, dedicated attempts have been made to search the Galactic bulge for extremely metal-poor stars. The bulge is known to be metal-rich, with a metallicity distribution function (MDF) peaking at [Fe/H]$\sim+0.3$ (e.g., \citealt{2013MNRAS.430..836N}; \citealt{2013A&A...552A.110G}). Furthermore, the huge number of stars in the bulge, the distance to the bulge, and the high degree of extinction in the Galactic plane make it practically very difficult to find metal-poor stars there. The ARGOS survey \citep{2013MNRAS.430..836N} spectroscopically studied 28,000 stars in the bulge at R$\approx10,000$, identifying only 16 stars with $-2.8$$\le$[Fe/H]$\le$$-2.0$, outlining the extent of the problem of finding metal-poor stars in a metallicity unbiased survey of the bulge.  Similar results were found in the BRAVA survey \citep{2012AJ....143...57K}. The APOGEE survey found five new metal-poor stars \citep{2013ApJ...767L...9G} with $-2.1$$\le$[Fe/H]$\le$$-1.6$ from $2400$ observed bulge stars. To date, no bulge star with [Fe/H]$<$$-2.1$ has been exposed to a high-resolution abundance analysis.
\\
This letter is the first in a series of papers exploring the results of the EMBLA\footnote{In Nordic mythology, Embla was the first woman, born in the middle of the world from the remains of giants.} (Extremely Metal-poor BuLge stars with AAOmega) survey, which aims to find the most metal-poor stars in the bulge.  Here we present the results of our initial observations, from which we have analysed four bulge stars with [Fe/H]$<-2$.
\section{Observations}
\begin{figure}
  \centering
  \includegraphics[width=0.99\columnwidth]{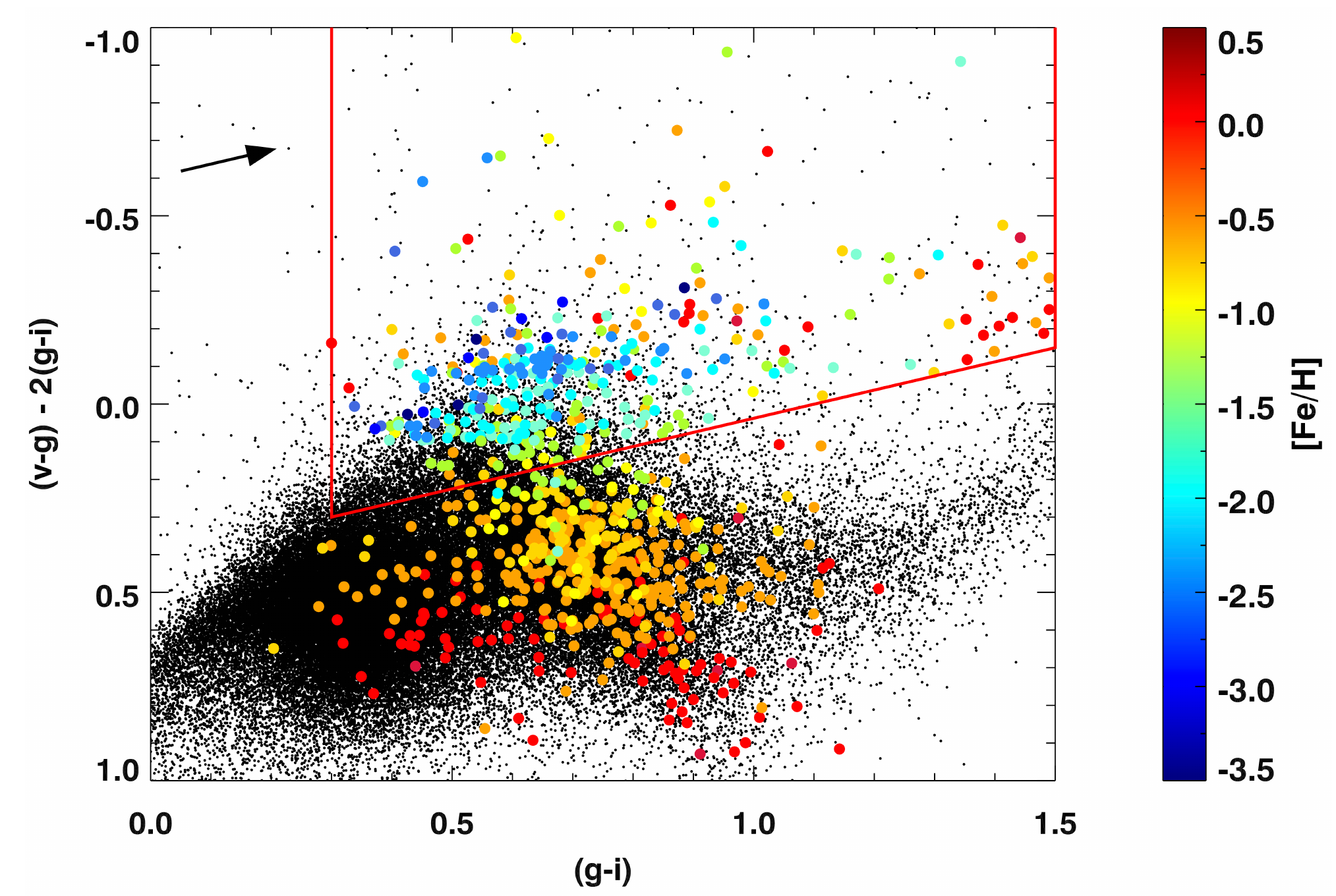}
  \caption{Two-colour plot using the $g$, $v$, and $i$ bands of SkyMapper to demonstrate the metallicity dependence on the $(v-g)-2(g-i)$ colour.  The coloured circles are data taken from both EMBLA and the ARGOS survey \citep{2013MNRAS.430..836N}, with [Fe/H] determined spectroscopically (field at (l,b)=(0,-10)).  The red trapezium shows our selection criteria for metal-poor candidates.  The arrow represents the mean reddening vector in this field, $E(B-V)$=0.17 \citep{1998ApJ...500..525S}.}
  \label{fig:photometry}
\end{figure}

\begin{figure}
  \includegraphics[width=0.99\columnwidth]{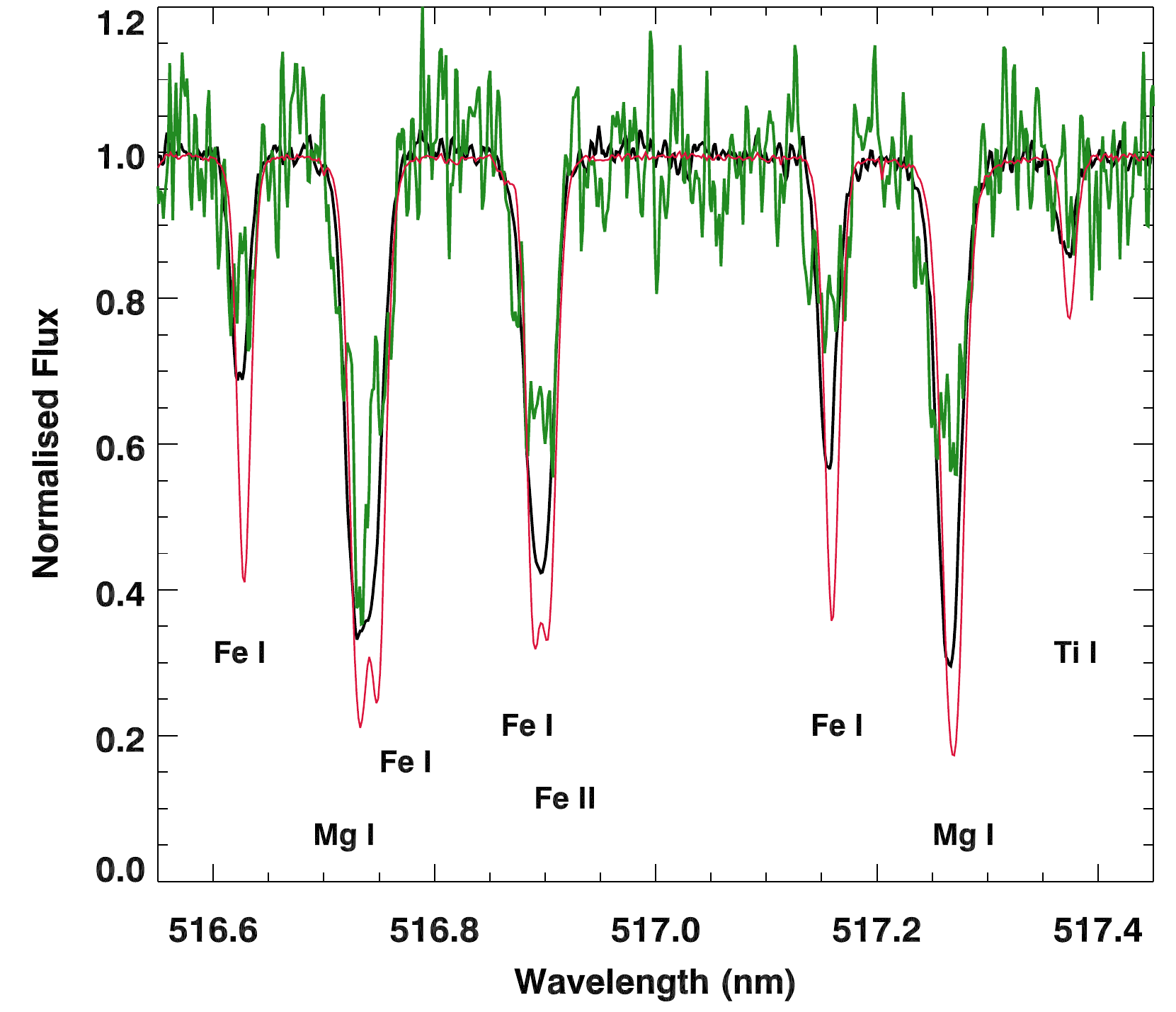}
  \caption{Comparison of the the FLAMES/UVES spectrum of Bulge-1 (black) and Bulge-3 (green), with the Gaia benchmark metal-poor halo giant \citep{2014A&A...564A.133J} HD 122563 (red), over the Mg triplet wavelength region.}
  \label{fig:spectrum}
\end{figure}

Our observations are conducted in three stages. We first acquire $uvgriz$ photometry from the SkyMapper telescope, identifying metal-poor candidates that we then confirm spectroscopically with the AAOmega multi-object spectrograph. Finally we analyse the most interesting stars using high-resolution spectra obtained with 8m class telescopes.
\\
The SkyMapper telescope is a 1.3m telescope capable of imaging in six bandpasses with a 5.7-square-degree field of view \citep{2007PASA...24....1K}. The filters have been designed to optimise both stellar and extragalactic astronomy; in particular the narrow $v$-band filter centred on the Ca\textsc{ii} K line provides a useful stellar metallicity indicator. We have obtained SkyMapper photometry, taken during commissioning, for more than 100 deg$^{2}$ of the bulge, with each field containing on the order of $10^{6}$ stars, ranging from 12th to 18th magnitude. From $(v-g)-2(g-i)$, $(g-i)$ two-colour diagrams we are able to select the most metal-poor candidates (Figure \ref{fig:photometry}). Figure \ref{fig:photometry} shows our selection 'box', which accounts for the effects of reddening.
\\
The AAOmega spectrograph on the AAT provides spectra of $\approx$$350$ stars (plus $\approx$$50$ sky and guide fibres) simultaneously over a 2 degree field-of-view \citep{2006SPIE.6269E..14S}.  Approximately 8500 bulge stars were observed in 2012 and 2013. All of these observations were taken using the 1700D grating for the red arm, and the 580V grating for the blue arm, providing a resolving power of about 10,000 over the 845-900 nm and of 1,300 over the 370-580 nm region.  The data were reduced using 2dfdr, and analysed with the \texttt{sick} pipeline \citep{2014ApJ..C} to measure radial velocities, and to determine the stellar parameters ($T_{\textup{eff}}$, log $g$, [Fe/H], and [$\alpha$/Fe]). 
\begin{figure}
  \centering
  \includegraphics[width=0.99\columnwidth]{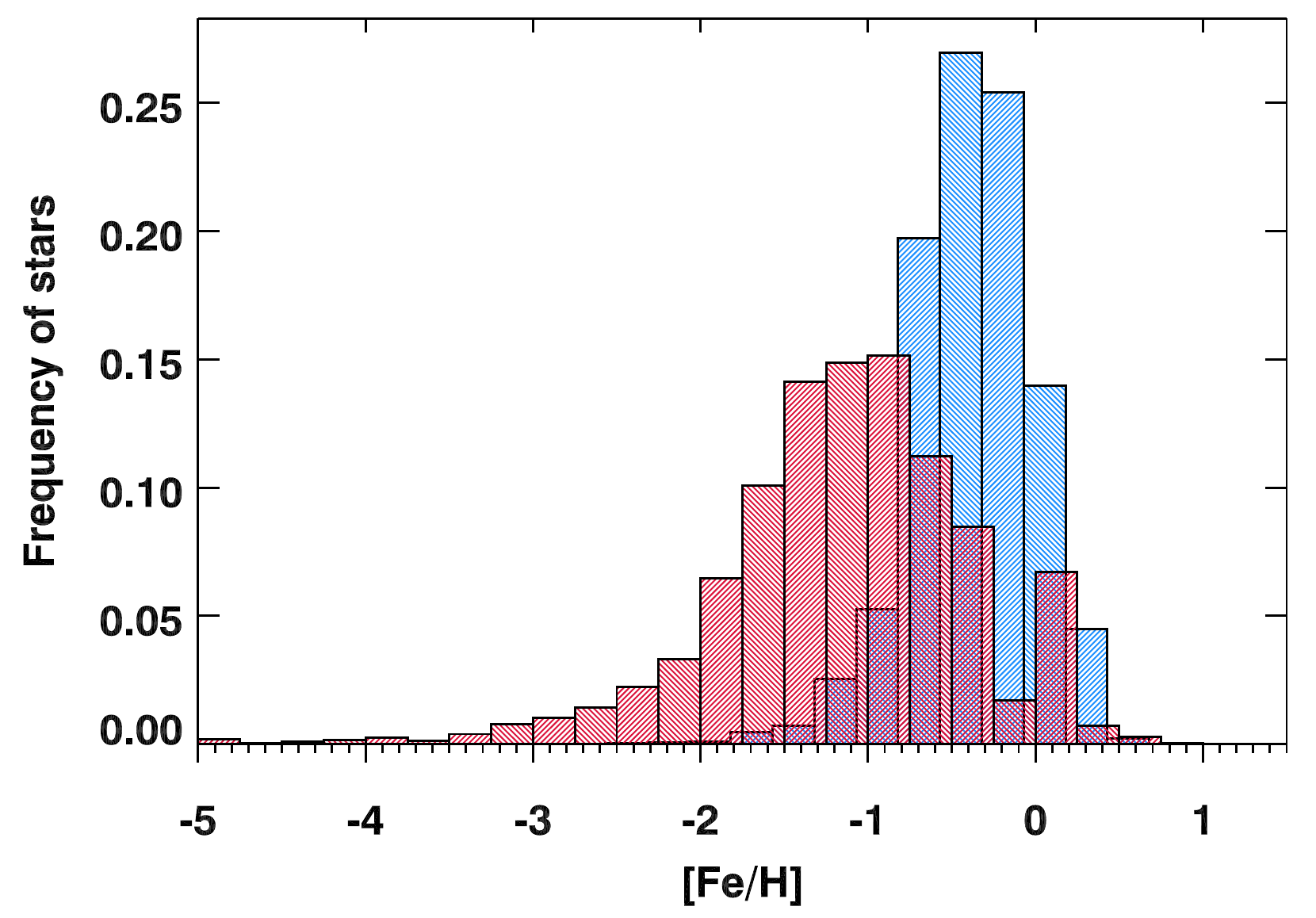}
  \caption{Raw metallicity distribution function, without accounting for selection effects, of the first $8,611$ spectra from the EMBLA survey (red), compared to the MDF of the ARGOS bulge survey (blue).  Both are normalised to have the same area.}
  \label{fig:histogram}
\end{figure}
\\
Figure \ref{fig:histogram} shows the raw MDF of the EMBLA survey, uncorrected for selection biases. In comparison to the relatively unbiased ARGOS MDF \citep{2013MNRAS.430..836N}, the average metallicity is approximately 0.8 dex lower with a significant tail of stars down into the extremely metal-poor regime. More than 300 stars have been found with [Fe/H]$<$$-2$. The full details of the EMBLA survey will be presented in future works.  
\\
From the first $3,600$ stars observed in April and July 2012, ten were immediately identified as very metal-poor candidates. Six of these targets were observed with FLAMES/UVES on the VLT \citep{2000SPIE.4008..534D} as part of the Gaia-ESO Survey (\citealt{2013Msngr.154...47R}) in May and August of 2012. The UVES observations have a resolving power of 47,000, using the 580nm setup. The data reduction of the FLAMES/UVES data in the survey is described in detail in \citet{2014A&A...565A.113S}.  Of these six stars, the signal-to-noise ratios of two were too poor to be able to gain any useful analysis from (S/N$\le8$). The other four had average S/N per pixel values ranging from 14 to 73, sufficient for the derivation of stellar parameters and chemical abundances. The spectra of two of the stars are shown in Figure \ref{fig:spectrum}, where they are compared to the Gaia-ESO benchmark metal-poor giant star, HD 122563 ([Fe/H]$=-2.64$, \citealt{2014A&A...564A.133J}), which has similar stellar parameters.  In addition, ten halo EMP candidates, similarly selected from SkyMapper photometry and intermediate resolution spectroscopy, were observed through Gaia-ESO.

\section{Stellar Parameters and Chemical Abundances}
\begin{table*}
\begin{minipage}{180mm}
\centering
\caption{Stellar parameters for the four bulge stars and ten halo stars observed as part of the Gaia-ESO Survey in May and August 2012. The SkyMapper naming convention is SMSS J(RA2000)+(Dec2000).}
\label{table:params}
\begin{tabular}{lrrrrrrrrr}
\hline
Star & l ($\degr$) & b ($\degr$) & $V_{GC}$  & d & S/N$^*$ & $T_{\textup{eff}}$ & log$(g)$ & [Fe/H]  & $\xi_{t}$ \\
 & & &(kms$^{-1}$)& (kpc) & & (K) & (cgs) & (dex) & (kms$^{-1}$) \\
\hline
SMSS J182153.85-341018.8 (Bulge-1) & 359.2 & -9.3 & -237.68 & 7.0 $\pm3.2$ & 73 & 4947 $\pm85$ & 1.41 $\pm0.49$ & -2.60 $\pm0.31$ & 2.3 $\pm0.2$\\
SMSS J183617.33-270005.3 (Bulge-2) & 7.1 & -8.9 & -129.48 & 5.3 $\pm1.9$ & 37 & 4926 $\pm137$ & 1.97 $\pm0.35$ & -2.72 $\pm0.28$ & 2.4 $\pm0.2$\\
SMSS J175510.50-412812.1 (Bulge-3) & 350.2 & -8.0 & -48.28 & 12.4 $\pm4.4$ & 14 & 5187 $\pm59$ & 2.23 $\pm0.31$ & -2.57 $\pm0.19$ & 2.0 $\pm0.2$ \\
SMSS J175652.43-413612.8 (Bulge-4) & 350.2 & -8.4 & 216.46 & 5.4 $\pm2.8$ & 14 & 5035 $\pm196$ & 2.65 $\pm0.54$ & -2.48 $\pm0.23$ & 1.5 $\pm0.2$\\
\\
SMSS J094755.04-102724.7 (Halo-1) & 246.8 & 31.8 & 309.16 & 2.1 $\pm1.1$ & 112 & 5258 $\pm195$ & 2.96 $\pm0.61$ & -2.80 $\pm0.17$ & 1.8 $\pm0.2$\\
SMSS J100915.77-412715.5 (Halo-2) & 273.1 & 11.8 & 383.89 & 3.6 $\pm1.8$ & 59 & 5266 $\pm200$ & 2.72 $\pm0.55$ & -2.42 $\pm0.18$ & 1.9 $\pm0.2$\\
SMSS J101427.85-405250.3 (Halo-3) & 273.6 & 12.9 & 284.67 & 2.0 $\pm0.9$ & 142 & 5136 $\pm123$ & 2.64 $\pm0.50$ & -1.99 $\pm0.20$ & 1.6 $\pm0.2$\\
SMSS J105806.38-154239.0 (Halo-4) & 266.9 & 39.1 & 519.06 & 8.0 $\pm2.6$ & 54 & 4907 $\pm58$ & 2.02 $\pm0.32$ & -2.39 $\pm0.12$ & 2.3 $\pm0.2$\\
SMSS J110053.36-132808.2 (Halo-5) & 266.0 & 41.3 & -73.40 & 9.4 $\pm4.6$ & 10 & 5194 $\pm125$ & 2.92 $\pm0.56$ & -2.30 $\pm0.40$ & 2.3 $\pm0.2$\\
SMSS J125551.46-450734.2 (Halo-6) & 303.8 & 17.7 & 496.02 & 4.3 $\pm3.1$ & 54 & 4957 $\pm215$ & 2.00 $\pm1.00$ & -2.62 $\pm0.19$ & 2.6 $\pm0.2$\\
SMSS J131358.51-460012.3 (Halo-7) & 307.0 & 16.7 & 284.04 & 8.5 $\pm3.2$ & 109 & 4744 $\pm62$ & 1.37 $\pm0.38$ & -2.34 $\pm0.22$ & 2.5 $\pm0.2$\\
SMSS J133013.60-434632.3 (Halo-8) & 310.3 & 18.5 & 314.65 & 11.2 $\pm3.6$ & 86 & 4558 $\pm54$ & 0.93 $\pm0.31$ & -2.42 $\pm0.08$ & 2.7 $\pm0.2$\\
SMSS J142148.60-440839.9 (Halo-9) & 319.6 & 15.8 & 232.93 & 12.5 $\pm4.8$ & 44 & 4761 $\pm65$ & 1.47 $\pm0.40$ & -2.42 $\pm0.18$ & 2.4 $\pm0.2$\\
SMSS J144100.90-400741.3 (Halo-10) & 324.7 & 18.1 & -138.96 & 6.9 $\pm2.7$ & 99 & 4921 $\pm93$ & 1.61 $\pm0.41$ & -2.45 $\pm0.17$ & 2.2 $\pm0.2$\\
\hline
%\caption{$^{a}$ Median S/N calculated across total wavelength range}
\end{tabular}
\end{minipage}\\
$^\ast$ Median S/N per pixel calculated across total wavelength range.
\end{table*}

The FLAMES/UVES spectra of the Gaia-ESO Survey are analysed by 13 different nodes \citep{2014arXiv1409.0568S}, each using the same MARCS model atmospheres \citep{2008A&A...486..951G} and line-lists (Heiter et al. in preparation), but different analysis techniques.
Due to the metal-poor nature of the stars, and with most pipelines being optimised for solar-metallicity stars, not all analysis nodes were able to establish robust parameters of the bulge and halo stars as well as the two metal-poor benchmark stars, HD 122563 (giant) and HD 140283 (subgiant). However three nodes provided accurate parameters for the majority of these metal-poor stars; the nodes IACAIP and Nice both used global fitting codes, while ULB used line-by-line analysis \citep{2014arXiv1409.0568S}.  In addition, we included the results of two further methods: firstly a modified version of the Lumba node pipeline used in the Smiljanic et al. analysis that uses the \textsc{sme} code \citep{1996A&AS..118..595V} for primarily H and Fe lines to determine stellar parameters, and secondly a similar analysis to that used in other SkyMapper EMP analyses (e.g., \citealt{2014Natur.506..463K}) using the \textsc{smh} code \citep{2014arXiv1405.5968C} in 1D LTE but with the Gaia-ESO line lists and atmospheres, and effective temperatures ($T_{\textup{eff}}$) measured from H lines instead.  The final parameters were evaluated by taking weighted averages of these five results.  The here derived stellar parameters, abundances and radial velocities have been adopted by the Gaia-ESO survey as the recommended values.  The uncertainties quoted are the calculated standard errors of the five parameter sets.
\\
Abundances for twelve elements were derived using the \textsc{smh} code (Table \ref{table:abund}). The uncertainties are formed from standard deviation of the line measurements taken in quadrature with the abundances differences due to stellar parameter uncertainties.  Some elements (Mg, Ca, Ti and Ni) were measured in all four bulge stars, however some of the elements could not be detected in the stars with lower S/N, while some elements were not detected at the lowest [Fe/H].  Additionally barium abundances were calculated from synthesis of the Ba lines rather than from equivalent widths, taking into account hyperfine splitting and isotopic shifts.

\section{Discussion}
\subsection{Bulge membership}
\begin{figure*}
\begin{minipage}{180mm}
\centering
  \includegraphics[width=0.75\columnwidth]{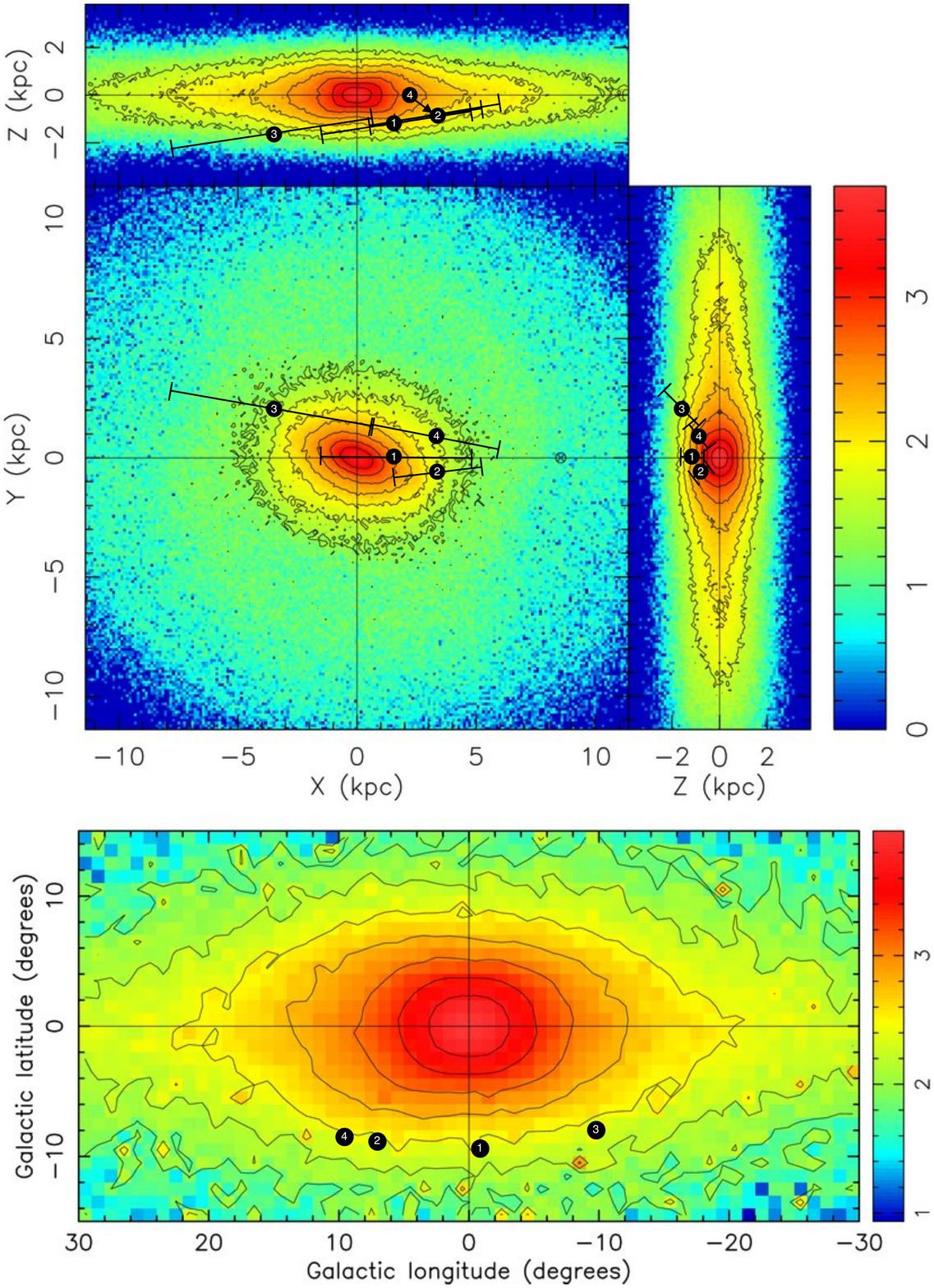}
  \caption{The four bulge stars plotted onto Figure 1 of \citealt{2010ApJ...720L..72S}, which shows the face-on and side-on views of the bulge constructed from their N-body model of the BRAVA survey data.  The position of the sun is marked with a cross on the x-axis.  Underneath, the position of the bulge stars as viewed from the Sun's location.}
  \label{fig:bulge}
\end{minipage}
\end{figure*}
All four stars were specifically taken from fields in the outer, southern part of the bulge ($-10\degr<l<8\degr$, $b\approx-8.5\degr$).  Distances have been estimated by calculating absolute luminosities based on our derived $T_{\textup{eff}}$, log$(g)$ and assuming $M_{*}=0.8M_{\sun}$, then fitting synthetic model fluxes from the $T_{\textup{eff}}$, log($g$), [Fe/H] and $E(B-V)$ \citep{1998ApJ...500..525S} of each star, from which the correction factor used to reconstruct the bolometric luminosities from 2MASS $JHK_{S}$ photometry is derived using the methodology described in \citet{2006MNRAS.373...13C} and \citet{2012ApJ...761...16C}.  The derived distances are given in Table \ref{table:params}.  Assuming a distance to the Galactic centre of 8.5 kpc and a bulge radius of 3 kpc, all but one are located inside the bulge.  When considering the more complex bar structure of the bulge, and given the large distance uncertainties, Bulge-3 is also consistent with residing in the bulge.  This is seen in Figure \ref{fig:bulge}, where the locations of all four stars have been projected onto an N-body model taken from \citealt{2010ApJ...720L..72S}.
\\
The radial velocities of our stars split them into two groups: those with velocities similar to that of the bulk of bulge stars, and those with larger velocities.  According to \citet{2013MNRAS.432.2092N}, the velocity dispersion of the bulge in the region of our stars is $\sigma=75.1$ km s$^{-1}$. Bulge-3 has a galactocentric velocity that would therefore be typical of a bulge star, but Bulge-1 and Bulge-4 have much larger velocities ($-237.68$ and $216.46$ km s$^{-1}$, respectively), and Bulge-2 lies in between the two groups.  These velocities are more characteristic of halo stars, and may indicate that although these stars are presently in the bulge, they are actually halo stars passing through. We intend to return to the important issue of kinematics for a much larger sample of stars in a future, detailed, analysis. For the time being we continue to refer to all four stars as bulge stars, given their location, but recognise that they may well have different origins from the typical bulge stars.

\subsection{Chemical composition}
\begin{figure}
  \centering
  \includegraphics[width=0.99\columnwidth]{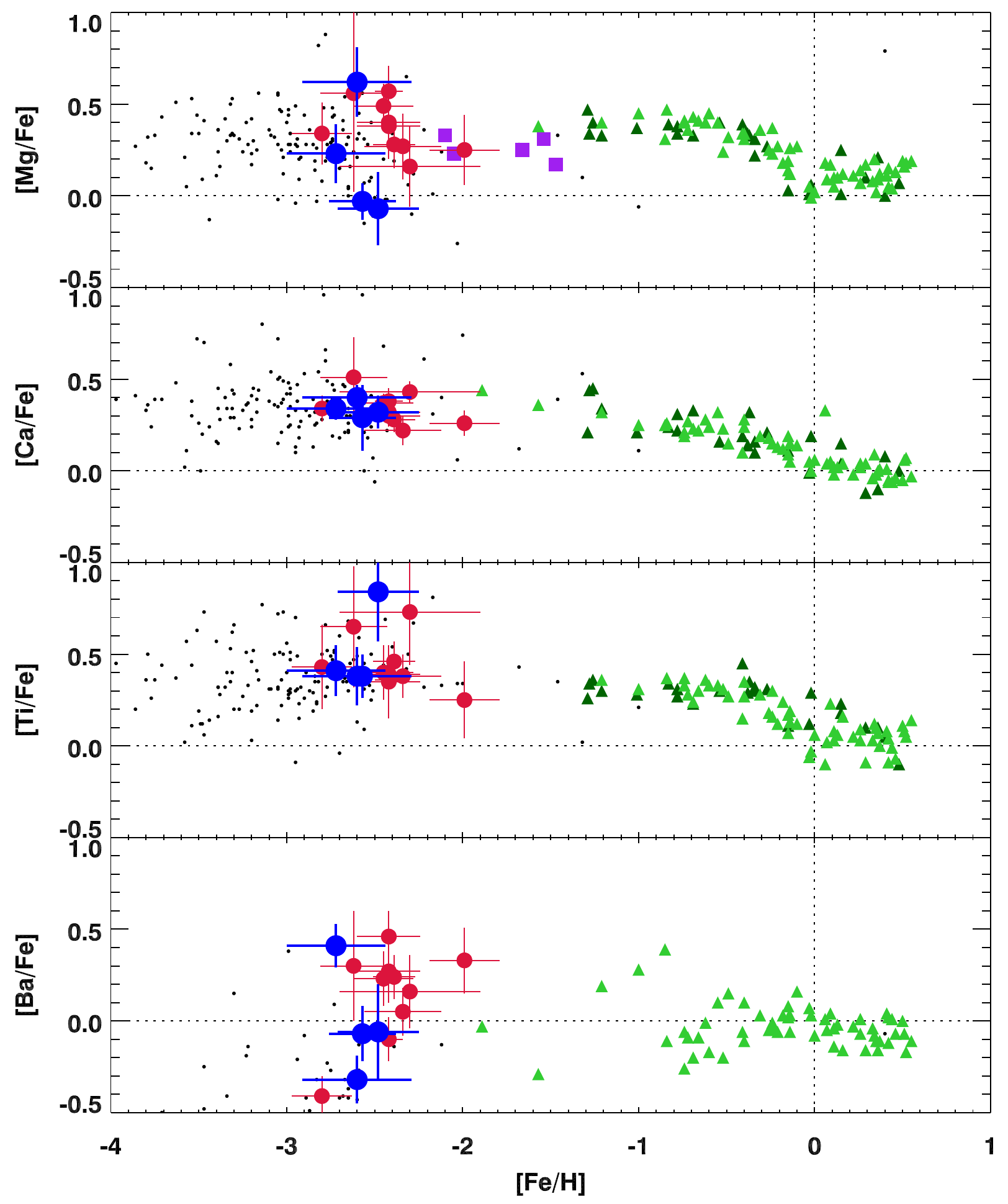}
  \caption{Comparison of the abundances in our bulge stars (blue circles), and the metal-poor halo stars from Gaia-ESO (red circles).  Also shown are bulge stars from \citet{2013A&A...549A.147B} (light green triangles), \citet{2010A&A...513A..35A} (dark green triangles), APOGEE (\citet{2013ApJ...767L...9G}, purple squares), as well as halo stars from \citet{2013ApJ...762...26Y} (black dots).}
  \label{fig:abundances}
\end{figure}

%The SkyMapper search for metal-poor stars has previously been shown to be successful in the halo, discovering the most iron-deficient star found to-date \citep{2014Natur.506..463K}. For the first time we have shown it is possible to search the crowded, metal-rich bulge and find the extremely rare metal-poor objects with a high success rate.  
All four program stars are confirmed (based on high-resolution spectroscopy) to have lower metallicities ([Fe/H]$<-2.4$) than any previously published metal-poor bulge star. We compare our four bulge stars in Figure \ref{fig:abundances} to metal-poor halo stars also identified by SkyMapper and observed as part of the Gaia-ESO Survey, as well as other published bulge and halo stars.  The ten halo stars also observed with the Gaia-ESO Survey have similar metallicities to the bulge stars, and all were analysed in an identical manner.
\begin{table}
%\begin{minipage}{180mm}
%\centering
\caption{Chemical abundances of the four bulge stars.}
\label{table:abund}
\begin{tabular}{lrrrr}
\hline
Element & Bulge-1 & Bulge-2 & Bulge-3 & Bulge-4 \\
\hline
$[$Na I/Fe$]$ &  0.65 $\pm0.23$ &  &  & 0.15 $\pm0.17$ \\
$[$Mg I/Fe$]$ & 0.62 $\pm0.19$ & 0.23 $\pm0.16$ & -0.03 $\pm0.10$ & -0.07 $\pm0.20$ \\
$[$Si I/Fe$]$ & 0.50 $\pm0.06$ & & & \\
$[$Ca I/Fe$]$ & 0.40 $\pm0.07$ & 0.24 $\pm0.06$ & 0.29 $\pm0.18$ & 0.32 $\pm0.09$ \\
$[$Sc II/Fe$]$ & 0.20 $\pm0.16$ & 0.22 $\pm0.15$ & & \\
$[$Ti II/Fe$]$ & 0.38 $\pm0.16$ & 0.41 $\pm0.14$ & 0.38 $\pm0.12$ & 0.84 $\pm0.27$ \\
$[$Cr I/Fe$]$ & -0.20 $\pm0.06$ & -0.27 $\pm0.06$& & -0.24 $\pm0.10$ \\
$[$Mn I/Fe$]$ & -0.50 $\pm0.04$ & & & \\
$[$Ni I/Fe$]$ & 0.02 $\pm0.03$ & 0.19 $\pm0.06$ & 0.11 $\pm0.25$ & 0.44 $\pm0.07$ \\
$[$Zn I/Fe$]$ & 0.45 $\pm0.06$ & & & \\
$[$Y II/Fe$]$ & -0.43 $\pm0.15$ & 0.34 $\pm0.14$ & & \\
$[$Ba II/Fe$]$ & -0.32 $\pm0.13$ & 0.41 $\pm0.12$ & -0.07 $\pm0.15$ & -0.06 $\pm0.26$ \\
\hline
\end{tabular}
%\end{minipage}
\end{table}
\\
Our abundance analysis reveals that all four bulge stars are significantly $\alpha$-enhanced (Fig. \ref{fig:abundances}), with average abundance ratios of [Mg/Fe]$=0.19$, [Si/Fe]$=0.50$ (one star), [Ca/Fe]$=0.34$, and [Ti/Fe]$=0.50$.  This enhancement is in line with the plateau shown in less metal-poor bulge stars of \citet{2013ApJ...767L...9G}, \citet{2010A&A...513A..35A}, and \citet{2013A&A...549A.147B}.  However, one obvious difference in this limited sample of metal-poor bulge stars is the intrinsic scatter in abundance, specifically in Mg and Ti.  Whereas the [Ca/Fe] ratios are all similar, matching the metal-poor halo stars and the more metal-rich bulge stars, for Mg and Ti the scatter is larger. Two stars have [Mg/Fe]$<0.0$, and one of those is very overabundant in Ti ([Ti/Fe]$=0.84$).  The scatter for these elements is comparable to that for the  halo stars analysed here, although in some cases the bulge stars may have larger scatter (noticeable in Mg, for example).  Compared to the larger sample of halo stars from \citet{2013ApJ...762...26Y}, given the limited statistics and remaining abundance uncertainties, our bulge and halo stars appear quite similar in [$\alpha$/Fe].
\\

Due to the wavelength region (480-680nm) and S/N of our spectra, it was only possible to measure Y and Ba in two of our bulge stars.  In these stars, both neutron-capture elements are under-abundant and similar to our sample of halo stars.  Again, the scatter is larger than for the more metal-rich bulge stars, although this is expected for neutron capture elements as seen in halo stars \citep{2007A&A...476..935F}.  There appear to be no obvious chemical differences in either neutron-capture or $\alpha$ abundances between those stars with very different velocities, although we caution that more stars are needed to confirm this.

\section{Conclusions}
Using the unique photometric capabilities of the SkyMapper telescope, and the large field-of-view, high multiplexing AAOmega spectrograph on the AAT, the EMBLA survey has already found more than 300 stars spectroscopically confirmed to have [Fe/H]$<$$-2.0$. We have presented an abundance analysis of four of these, observed in high-resolution as part of the Gaia-ESO Survey.  These four are all considerably more metal-poor than any previously studied bulge star ($-2.72$$<$[Fe/H]$<$$-2.48$) and are chemically similar to metal-poor halo stars at similar [Fe/H]. The four stars are the first of many which will be studied at high-resolution by the EMBLA survey using Magellan and VLT, which will allow us to study the metal-poor tail of the bulge's MDF, make a detailed comparison with the halo and study the oldest stars in the Universe, many of which would have formed at z$\approx$$15$.  

\section*{Acknowledgements}
We greatly acknowledge generous funding from the Australian Research Council (grants FL110100012 and DP120101237). T.B. was funded by grant No. 621-2009- 3911 from The Swedish Research Council. Support from the Swedish National Space Board is acknowledged. Based on data products from observations made with ESO Telescopes at the La Silla Paranal Observatory under programme ID 188.B-3002. This work was partly supported by the Eu- ropean Union FP7 programme through ERC grant num- ber 320360 and by the Leverhulme Trust through grant RPG-2012-541. We acknowledge the support from INAF and Ministero dell’ Istruzione, dell’ Universita` ’e della Ricerca (MIUR) in the form of the grant ”Premiale VLT 2012”. The results presented here benefit from discussions held during the Gaia-ESO workshops and conferences supported by the ESF (European Science Foundation) through the GREAT Research Network Programme.  MZ acknowledges support from Fondecyt Regular 1110393, the BASAL CATA PFB-06 and by the Chilean Ministry for the Economy, Development, and Tourism's Programa Iniciativa Cient\'\i fica Milenio through grant IC12009, awarded to the Millennium Institute of Astrophysics.

\bibliography{references}

\begin{thebibliography}{36}
\expandafter\ifx\csname natexlab\endcsname\relax\def\natexlab#1{#1}\fi

\bibitem[{Abel, Bryan \& Norman(2002)Abel, Bryan, \&
  Norman}]{2002Sci...295...93A}
Abel T., Bryan G., Norman M., 2002, Science, 295, 93

\bibitem[{Alves-Brito {et~al}\mbox{.}(2010)Alves-Brito, Mel{\'e}ndez, Asplund,
  Ram{\'\i}rez, \& Yong}]{2010A&A...513A..35A}
Alves-Brito A., Mel{\'e}ndez J., Asplund M., Ram{\'\i}rez I., Yong D., 2010,
  A{\&}A, 513, 35

\bibitem[{Bensby {et~al}\mbox{.}(2013)Bensby, Yee, Feltzing, Johnson, Gould,
  Cohen, Asplund, Mel{\'e}ndez, Lucatello, Han, Thompson, Gal-Yam, Udalski,
  Bennett, Bond, Kohei, Sumi, Suzuki, Suzuki, Takino, Tristram, Yamai, \&
  Yonehara}]{2013A&A...549A.147B}
Bensby T. {et~al.}, 2013, A{\&}A, 549, 147

\bibitem[{Bromm \& Loeb(2006)}]{2006ApJ...642..382B}
Bromm V., Loeb A., 2006, ApJ, 642, 382

\bibitem[{Casagrande, Portinari \& Flynn(2006)Casagrande, Portinari, \&
  Flynn}]{2006MNRAS.373...13C}
Casagrande L., Portinari L., Flynn C., 2006, MNRAS, 373, 13

\bibitem[{Casagrande {et~al}\mbox{.}(2012)Casagrande, Ram{\'\i}rez,
  Mel{\'e}ndez, \& Asplund}]{2012ApJ...761...16C}
Casagrande L., Ram{\'\i}rez I., Mel{\'e}ndez J., Asplund M., 2012, ApJ, 761, 16

\bibitem[{{Casey}(2014)}]{2014arXiv1405.5968C}
{Casey} A.~R., 2014, arXiv:1405.5968

\bibitem[{Casey(2014b)}]{2014ApJ..C}
Casey A.~R., 2014b, ApJS, submitted

\bibitem[{Christlieb {et~al}\mbox{.}(2008)Christlieb, Sch{\"o}rck, Frebel,
  Beers, Wisotzki, \& Reimers}]{2008A&A...484..721C}
Christlieb N., Sch{\"o}rck T., Frebel A., Beers T.~C., Wisotzki L., Reimers D.,
  2008, A{\&}A, 484, 721

\bibitem[{Clark {et~al}\mbox{.}(2011)Clark, Glover, Smith, Greif, Klessen, \&
  Bromm}]{2011Sci...331.1040C}
Clark P.~C., Glover S. C.~O., Smith R.~J., Greif T.~H., Klessen R.~S., Bromm
  V., 2011, Science, 331, 1040

\bibitem[{Cooke, Pettini \& Murphy(2012)Cooke, Pettini, \&
  Murphy}]{2012MNRAS.425..347C}
Cooke R., Pettini M., Murphy M.~T., 2012, Monthly Notices of the Royal
  Astronomical Society, 425, 347

\bibitem[{Dekker {et~al}\mbox{.}(2000)Dekker, D'Odorico, Kaufer, Delabre, \&
  Kotzlowski}]{2000SPIE.4008..534D}
Dekker H., D'Odorico S., Kaufer A., Delabre B., Kotzlowski H., 2000, SPIE,
  4008, 534

\bibitem[{Diemand, Madau \& Moore(2005)Diemand, Madau, \&
  Moore}]{2005MNRAS.364..367D}
Diemand J., Madau P., Moore B., 2005, MNRAS, 364, 367

\bibitem[{Francois {et~al}\mbox{.}(2007)Francois, Depagne, Hill, Spite, Spite,
  Plez, Beers, Andersen, James, Barbuy, Cayrel, Bonifacio, Molaro,
  Nordstr{\"o}m, \& Primas}]{2007A&A...476..935F}
Francois P. {et~al.}, 2007, A{\&}A, 476, 935

\bibitem[{Frebel, Kirby \& Simon(2010)Frebel, Kirby, \&
  Simon}]{2010Natur.464...72F}
Frebel A., Kirby E.~N., Simon J.~D., 2010, Nature, 464, 72

\bibitem[{Frebel \& Norris(2013)}]{2013pss5.book...55F}
Frebel A., Norris J.~E., 2013, Planets, Stars and Stellar Systems Vol. 5, p. 55

\bibitem[{Garc{\'\i}a~P{\'e}rez {et~al}\mbox{.}(2013)Garc{\'\i}a~P{\'e}rez,
  Cunha, Shetrone, Majewski, Johnson, Smith, Schiavon, Holtzman, Nidever,
  Zasowski, Allende~Prieto, Beers, Bizyaev, Ebelke, Eisenstein, Frinchaboy,
  Girardi, Hearty, Malanushenko, Malanushenko, Meszaros, O'Connell, Oravetz,
  Pan, Robin, Schneider, Schultheis, Skrutskie, Simmonsand, \&
  Wilson}]{2013ApJ...767L...9G}
Garc{\'\i}a~P{\'e}rez A.~E. {et~al.}, 2013, ApJL, 767, L9

\bibitem[{Gonzalez {et~al}\mbox{.}(2013)Gonzalez, Rejkuba, Zoccali, Valent,
  Minniti, \& Tobar}]{2013A&A...552A.110G}
Gonzalez O.~A., Rejkuba M., Zoccali M., Valent E., Minniti D., Tobar R., 2013,
  A{\&}A, 552, 110

\bibitem[{Gustafsson {et~al}\mbox{.}(2008)Gustafsson, Edvardsson, Eriksson,
  J{\o}rgensen, Nordlund, \& Plez}]{2008A&A...486..951G}
Gustafsson B., Edvardsson B., Eriksson K., J{\o}rgensen U.~G., Nordlund {\AA}.,
  Plez B., 2008, A{\&}A, 486, 951

\bibitem[{Jofre {et~al}\mbox{.}(2014)Jofre, Heiter, Soubiran, Blanco-Cuaresma,
  Worley, Pancino, Cantat-Gaudin, Magrini, Bergemann,
  Gonz{\'a}lez~Hern{\'a}ndez, Hill, Lardo, de~Laverny, Lind, Masseron, Montes,
  Mucciarelli, Nordlander, Recio-Blanco, Sobeck, Sordo, Sousa, Tabernero,
  Vallenari, \& Van~Eck}]{2014A&A...564A.133J}
Jofre P. {et~al.}, 2014, A{\&}A, 564, 133

\bibitem[{Keller {et~al}\mbox{.}(2014)Keller, Bessell, Frebel, Casey, Asplund,
  Jacobson, Lind, Norris, Yong, Heger, Magic, Da~Costa, Schmidt, \&
  Tisserand}]{2014Natur.506..463K}
Keller S.~C. {et~al.}, 2014, Nature, 506, 463

\bibitem[{Keller {et~al}\mbox{.}(2007)Keller, Schmidt, Bessell, Conroy,
  Francis, Granlund, Kowald, Oates, Martin-Jones, Preston, Tisserand,
  Vaccarella, \& Waterson}]{2007PASA...24....1K}
Keller S.~C. {et~al.}, 2007, PASA, 24, 1

\bibitem[{Kunder {et~al}\mbox{.}(2012)Kunder, Koch, Rich, de~Propris, Howard,
  Stubbs, Johnson, Shen, Wang, Robin, Kormendy, Soto, Frinchaboy, Reitzel,
  Zhao, \& Origlia}]{2012AJ....143...57K}
Kunder A. {et~al.}, 2012, The Astronomical Journal, 143, 57

\bibitem[{Nakamura \& Umemura(2001)}]{2001ApJ...548...19N}
Nakamura F., Umemura M., 2001, ApJ, 19

\bibitem[{Ness {et~al}\mbox{.}(2013{\natexlab{a}})Ness, Freeman, Athanassoula,
  Wylie-De-Boer, Bland~Hawthorn, Asplund, Lewis, Yong, Lane, \&
  Kiss}]{2013MNRAS.430..836N}
Ness M. {et~al.}, 2013{\natexlab{a}}, MNRAS, 430, 836

\bibitem[{Ness {et~al}\mbox{.}(2013{\natexlab{b}})Ness, Freeman, Athanassoula,
  Wylie-de Boer, Bland-Hawthorn, Asplund, Lewis, Yong, Lane, Kiss, \&
  Ibata}]{2013MNRAS.432.2092N}
Ness M. {et~al.}, 2013{\natexlab{b}}, MNRAS, 432, 2092

\bibitem[{Norris {et~al}\mbox{.}(2013)Norris, Bessell, Yong, Christlieb,
  Barklem, Asplund, Murphy, Beers, Frebel, \& Ryan}]{2013ApJ...762...25N}
Norris J.~E. {et~al.}, 2013, ApJ, 762, 25

\bibitem[{Randich \& Gilmore(2013)}]{2013Msngr.154...47R}
Randich S., Gilmore G., 2013, ESO Messenger, 154, 47

\bibitem[{Sacco {et~al}\mbox{.}(2014)Sacco, Morbidelli, Franciosini, Maiorca,
  Randich, Modigliani, Gilmore, Asplund, Binney, Bonifacio, Drew, Feltzing,
  Ferguson, Jeffries, Micela, Negueruela, Prusti, Rix, Vallenari, Alfaro,
  Allende~Prieto, Babusiaux, Bensby, Blomme, Bragaglia, Flaccomio, Francois,
  Hambly, Irwin, Koposov, Korn, Lanzafame, Pancino, Recio-Blanco, Smiljanic,
  Van~Eck, Walton, Bergemann, Costado, de~Laverny, Heiter, Hill, Hourihane,
  Jackson, Jofre, Lewis, Lind, Lardo, Magrini, Masseron, Prisinzano, \&
  Worley}]{2014A&A...565A.113S}
Sacco G.~G. {et~al.}, 2014, A{\&}A, 565, 113

\bibitem[{Schlegel, Finkbeiner \& Davis(1998)Schlegel, Finkbeiner, \&
  Davis}]{1998ApJ...500..525S}
Schlegel D., Finkbeiner D., Davis M., 1998, ApJ, 500, 525

\bibitem[{Sharp {et~al}\mbox{.}(2006)Sharp, Saunders, Smith, Churilov, Correll,
  Dawson, Farrel, Frost, Haynes, Heald, Lankshear, Mayfield, Waller, \&
  Whittard}]{2006SPIE.6269E..14S}
Sharp R. {et~al.}, 2006, SPIE, 6269, 14

\bibitem[{Shen {et~al}\mbox{.}(2010)Shen, Rich, Kormendy, Howard, de~Propris,
  \& Kunder}]{2010ApJ...720L..72S}
Shen J., Rich R.~M., Kormendy J., Howard C.~D., de~Propris R., Kunder A., 2010,
  ApJL, 720, L72

\bibitem[{Smiljanic {et~al}\mbox{.}(2014)Smiljanic, Korn, Bergemann, Frasca,
  Magrini, Masseron, Pancino, Ruchti, San~Roman, Sbordone, Sousa, Tabernero,
  Tautvaisiene, Valentini, Weber, Worley, Adibekyan, Allende~Prieto,
  Barisevicius, Biazzo, Blanco-Cuaresma, Bonifacio, Bragaglia, Caffau,
  Cantat-Gaudin, Chorniy, de~Laverny, Delgado-Mena, Donati, Duffau,
  Franciosini, Friel, Geisler, Gonz{\'a}lez~Hern{\'a}ndez, Gruyters, Guiglion,
  Hansen, Heiter, Hill, Jacobson, Jofre, Jonsson, Lanzafame, Lardo, Ludwig,
  Maiorca, Mikolaitis, Montes, Morel, Mucciarelli, Munoz, Nordlander, Pasquini,
  Puzeras, Recio-Blanco, Ryde, Sacco, Santos, Serenelli, Sordo, Soubiran,
  Spina, Steffen, Vallenari, Van~Eck, Villanova, Gilmore, Randich, Asplund,
  Binney, Drew, Feltzing, Ferguson, Jeffries, Micela, Negueruela, Prusti, Rix,
  Alfaro, Babusiaux, Bensby, Blomme, Flaccomio, Francois, Irwin, Koposov,
  Walton, Bayo, Carraro, Costado, Damiani, Edvardsson, Hourihane, Jackson,
  Lewis, Lind, Marconi, Martayan, Monaco, Morbidelli, Prisinzano, \&
  Zaggia}]{2014arXiv1409.0568S}
Smiljanic R. {et~al.}, 2014, arXiv:1409.0568

\bibitem[{Tumlinson(2010)}]{2010ApJ...708.1398T}
Tumlinson J., 2010, ApJ, 708, 1398

\bibitem[{Valenti \& Piskunov(1996)}]{1996A&AS..118..595V}
Valenti J.~A., Piskunov N., 1996, A{\&}AS, 118, 595

\bibitem[{Yong {et~al}\mbox{.}(2013)Yong, Norris, Bessell, Christlieb, Asplund,
  Beers, Barklem, Frebel, \& Ryan}]{2013ApJ...762...26Y}
Yong D. {et~al.}, 2013, ApJ, 762, 26

\end{thebibliography}
\bibliographystyle{mn2e2}

\end{document}